\documentclass[twocolumn,twoside,slac_two]{revtex4}
\usepackage{graphicx}
\usepackage{fancyhdr}
\begin{document}

\title{Searching for Unmodeled Sources Using the Earth Occultation Data from the \emph{Fermi} GBM}

%

\author{James Rodi, Gary L. Case, Michael L. Cherry}
\affiliation{Department of Physics \& Astronomy, Louisiana State University, Baton Rouge, LA 70803, USA}

\author{Ascension Camero-Arranz, Mark H. Finger}
\affiliation{Universities Space Research Association, Huntsville, AL}

\author{Peter Jenke, Colleen A. Wilson-Hodge}
\affiliation{NASA Marshall Space Flight Center, Huntsville, AL}

\author{Vandiver Chaplin}
\affiliation{University of Alabama in Huntsville, Huntsville, AL}

\begin{abstract}
Employing the 12 NaI detectors in the \emph{Fermi} GBM, the Earth Occultation Technique (EOT) can be used to measure the fluxes of x-ray and gamma-ray sources.  Each time a source passes behind the Earth (or emerges from behind the Earth), a step-like feature is produced in the detector count rate.  With a predefined catalog of source positions, the times of the occultation steps can be calculated, the individual steps fit, and the fluxes derived.  However, in order to find new sources and generate a complete catalog, a method is needed for generating an image of the sky.  An imaging algorithm has been developed to generate all-sky images using the GBM data.  Here we present imaging results from \( \sim 2.5 \) years of data in the 12-25 keV and 100-300 keV energy bands.

\end{abstract}

\maketitle

\thispagestyle{fancy}


\section{Introduction}
The hard x-ray/soft gamma-ray sky is extremely variable, and so the ability to continually monitor the sky at these energies is important.  Especially above \( \sim \)200 keV, imaging with the wide-field coded aperture technique is difficult ~\cite{bat-paper} so other detection techniques are required.  With these two considerations in mind, the Earth Occultation Technique (EOT) ~\cite{exampl-ref-2,exampl-ref-3} applied 12 NaI detectors from the Gamma-ray Burst Monitor (GBM) onboard \emph{Fermi} is used to provide a the large field of view for continuous observations of the sky at energies 12-1000 keV.  EOT has been used successfully to study sources over this range of energies with the Burst and Transient Source Experiment (BATSE) on the Compton Gamma Ray Observatory (CGRO) ~\cite{exampl-ref-2,exampl-ref-3} and with GBM ~\cite{exampl-ref-5,exampl-ref-1,exampl-ref-6}.  Because EOT requires an input catalog of source positions to monitor sources, a new imaging approach is employed to discover sources not in the catalog.  The imaging algorithm that has been developed is a tomographic method that uses projections of the sky over the course of the orbital precession period of the satellite ( \( \sim 53 \) days) to generate an image.  Results presented here include all-sky images for \( \sim 2.5 \) years of GBM data for the 12-25 keV and 100-300 keV broad CTIME energy bands and a table of sources that have been added to the EOT input catalog based on the imaging results.  

\section{Earth Occultation Technique}
As demonstrated with BATSE, monitoring known hard x-ray/soft gamma-ray sources is possible with simple non-imaging detectors with EOT.  The count rate in a detector viewing the source decreases like a step function every time a source is occulted by the Earth.  A similar feature, but with opposite polarity, occurs in the detector count rates when the source comes out of occultation.  Occultation times can be predicted using the coordinates of the source and the spacecraft positions.  The occultation time is defined to be the time at which the transmission through the atmosphere is 50\% at 100 keV.  In calculating a flux for the source of interest, a 4-minute window of 0.256-sec CTIME count-rate data, centered on the calculated occultation time, is fit to a quadratic background plus a source model for each source that occults during the window.  The source model is constructed from combining an assumed source spectrum with an atmospheric transmission model convolved with the detector response.  For each energy channel, a scale factor is derived from the fit to the data and averaged over the detectors viewing the source.  The weighted average scale factor for the source of interest is multiplied by the model flux to determine the final flux ~\cite{exampl-ref-5,exampl-ref-1}.

\section{Earth Occultation Imaging}
Since EOT requires an input catalog with predetermined source positions, another method is needed to search for sources absent from the input catalog.  An imaging method conceptually similar to x-ray computed  tomography used in medical imaging has been developed.  When a source occults, the Earth's limb can be projected onto the sky resulting in a ``slice'' on the sky over which counts are integrated.  During the orbital precession of the spacecraft (\( \sim 53 \) days for \emph{Fermi}), the angle between the source position and the plane of the satellite orbit changes.  Thus over the course of an orbital precession period a range of projections are sampled.  The projections from rising and setting occultation steps can be combined to generate an image of the sky and localize a source.  The angular resolution of this method is limited by the finite time required to go from full transmission to full attenuation for a source.  The duration of a step for a source occulting in the plane of the satellite's orbit is \( \sim 8\) seconds.  As the angle between the orbital plane and the source increases, the duration of the occultation step increases by \( \sim 1/ \cos( \beta ) \), where \( \beta \) is the elevation angle between the source and the orbital plane of the satellite.  The angular resolution is \( 360^{\circ} \times \Delta t / P \), where \( \Delta t \) is the occultation duration and \( P \) is the satellite's orbital period ( \( \sim 96 \) minutes).  The angular resolution ranges from \( \sim 0.5^{\circ} \text{ at } \beta = 0^{\circ} \text{ to } \sim 1.25^{\circ} \text{ at } \beta = 66^{\circ} \).  When \( \beta > 66^{ \circ} \), occultation no longer occur.  

Instead of a catalog of known source positions, our imaging technique starts with a catalog of virtual sources covering te sky with spacing of \( \sim 0.25^{\circ} \), resulting in a catalog of \( \sim \) 660,000 positions on the sky.  With these virtual source positions and the spacecraft position history, occultation times can be predicted.  A 4-minute window of GBM CTIME counts data centered on each occultation time is selected for the detectors that view the step at less that \( 75^{\circ} \) from the detector normal.  A first-order response correction is applied to the data from these detectors.  To be able to sum rise and set steps together, occultation windows for rise steps are converted to set steps by rotating the data about the occultation time at the center of the window.  A weighted average of all the windows for a virtual source is calculated for an orbital precession period.  The averaged window is rebinned from the standard CTIME resolution of 0.256 seconds to 2.048-second resolution, then passed through a differential filter of the form

\begin{displaymath}
o_i = \frac{ \sum_{j=i+f_a}^{j=i+f_a+f_b} r_j - \sum_{j=i-f_a}^{j=i-f_a-f_b} r_j}{f_b},
\end{displaymath}
where \( r_j \) is the number of counts in bin \( j \), \( f_a \) is the inner boundary of the filter, and \( f_b \) is the outer boundary ~\cite{exampl-ref-4}.  The values of \( f_a \text{ and } f_b \) used here are 3 bins and 8 bins, respectively.  The differential filter generates a dip at the occultation time while smoothing background features, where the amplitude of the dip is related to the intensity of the source.  To calculate the amplitude, the window is fit in two sections.  Bins within \( \pm (2 f_a + f_b ) \) of the occultation time are fit to a polynomial.  The outer portions of the window are fit with a spline function and joined by a straight line.  The amplitude of the virtual source is found by taking the difference between the two fits at the occultation time.  

\begin{figure}[t]
\includegraphics[angle=90,scale=0.37, trim=10mm 25mm 0mm 25mm,clip=true]{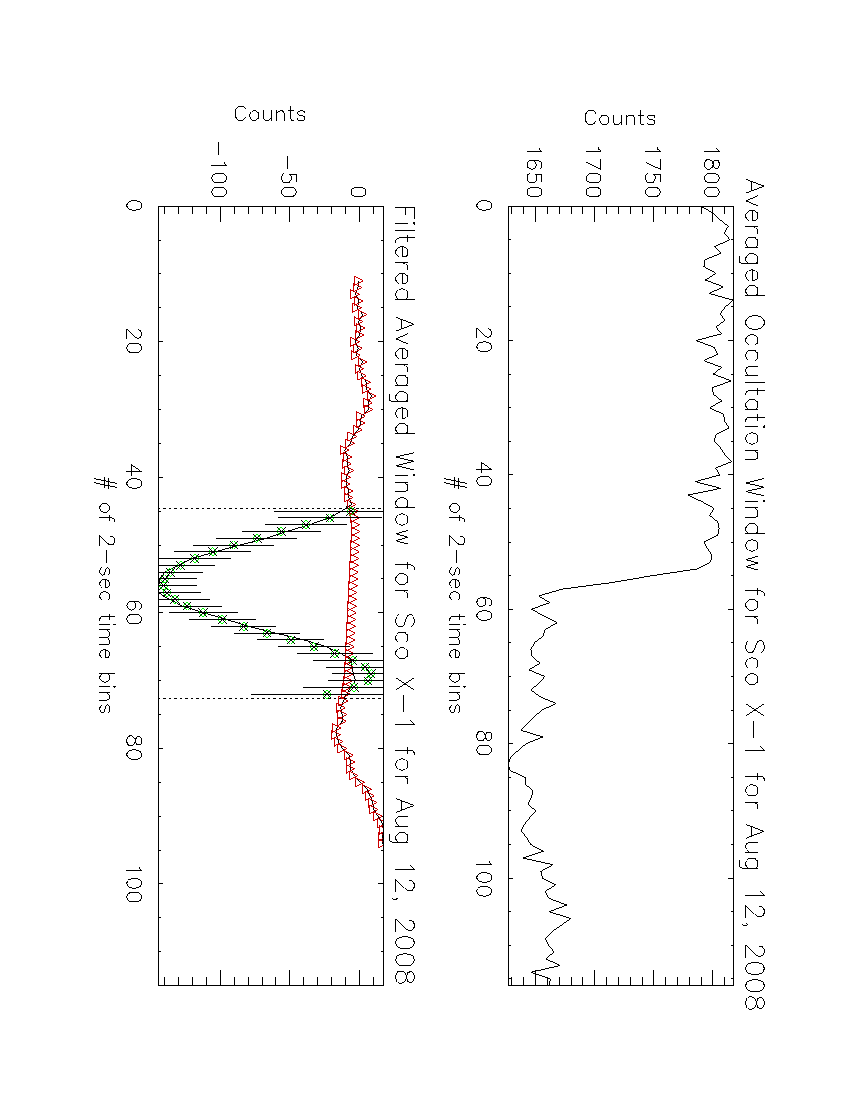}
\vspace{-12mm}
\caption{\emph{Top}: Averaged window for 1 day for Sco X-1.  \emph{Bottom}: The filtered data with the central portion fit to a polynomial (green), and the outer background portion fit to a spline function (red). }
\vspace{-5mm}
\end{figure}

The top panel in Figure 1 shows the averaged occultation window for one day for a point at the location of Sco X-1.  The bottom panel shows the filtered window with the polynomial fit in green and the spline fit in red.  The dashed lines are the boundaries for the polynomial fit.

\subsection{Systematic Effects}
Four systematic effects are explicitly taken into account in the processing: 1) high-spin rate data, 2) South Atlantic Analomy (SAA) passes, 3) source confusion with bright sources, and 4) virtual sources in which the rise and set steps of a bright source dominate the spline background fit.  The first two effects are removed by making cuts on the data to remove times when those issues are present.  High spin rate effects occur when the spacecraft slews, which leads to a rapidly varying background during the 4-minute window.  The SAA pass effect comes from activation in the detectors during passes through the SAA and results in large spikes in the data, which again leads to a rapidly varying background.  

The last two systematic effects are caused by bright sources.  The issue of source confusion arises from the ambiguity of where along the Earth's limb the measured flux originates.  Since the GBM detectors have no direct imaging capability, the flux for an occultation time can be attributed to anywhere on the sky along the Earth's limb.  Consequently, bright sources (e.g. Crab, Cyg X-1, Sco X-1) cause ``X'' patterns that can extend for tens of degrees.  These ``arms'' make it difficult to determine the presence of fainter sources as they overwhelm any flux that comes from possible faint sources.  To mitigate the effects of source confusion, an algorithm has been developed to ignore occultation windows for virtual sources when bright sources occult close in time ( \( < 11 \) seconds) but far away on the sky.  

The final systematic effect involves virtual sources at roughly the same declination of a bright source and a few degrees away in right ascension.  The results of the fits are large negative amplitudes for virtual sources.  The virtual source is close enough on the sky to the bright source so that both rise and set occultation steps for the bright source occur inside the 4-minute window, but the two sources are far enough away such that when the window is filtered, the rise steps and the set steps of the bright source remain separated.  Consequently, the steps from the bright source are treated as the background level, which causes the calculated amplitude of the virtual source to be extremely negative as the background level is larger then the source at the occultation time.  To reduce this issue the fit boundaries are shifted towards the occultation time at the center of the window and reduce the effect of the bright source on the background fit.  

Figure 2 demonstrates this problem by using a source close to the Crab (\( \alpha = 83.63^{\circ} \text{, } \delta = 22.02^{\circ} \)) at \( \alpha = 80.75^{\circ} \text {, }   \delta = 22.0^{\circ} \) for a precession period.  The top panel in Figure 2 shows the rise and set occultation steps for the Crab, roughly at bin 35 and bin 75.  The bottom panel shows how the filtered Crab steps misrepresent the background and thus the calculated amplitude of the virtual source.  

\begin{figure}[h]
\vspace{-3mm}
\includegraphics[angle=90,scale=0.35, trim=0mm 25mm 10mm 25mm,clip=true]{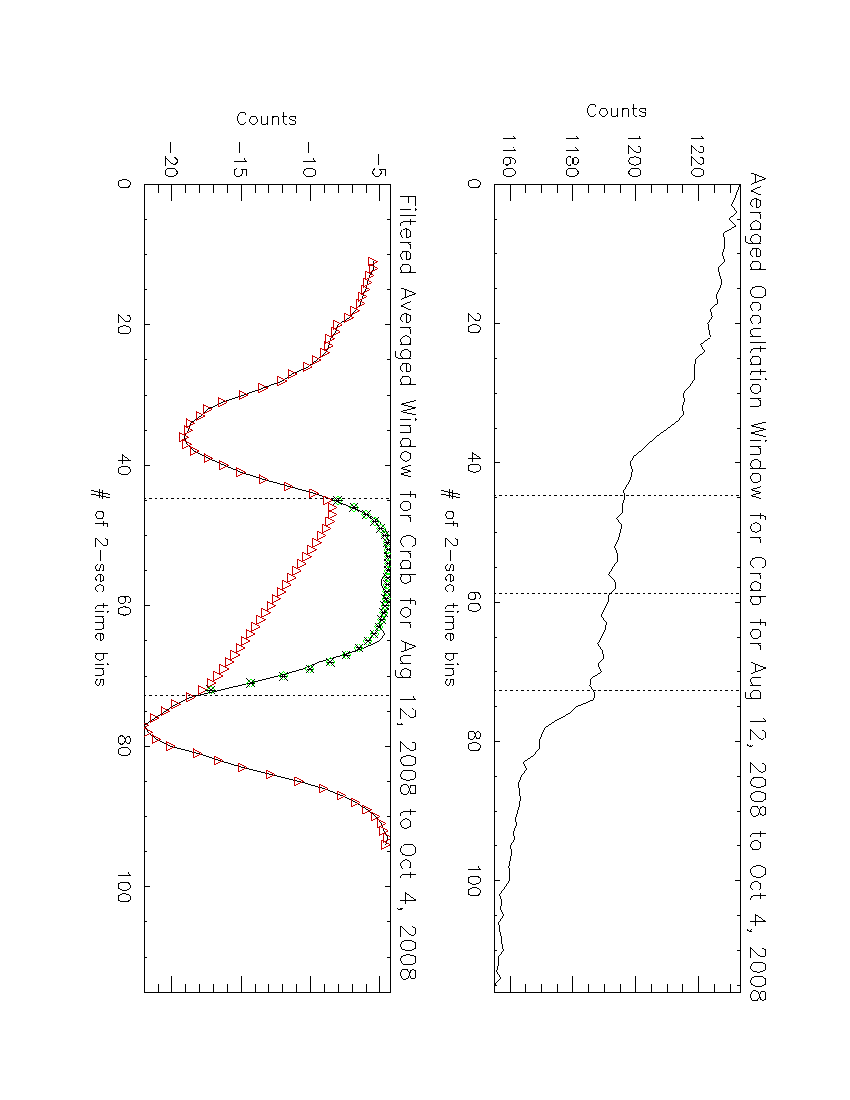}
\vspace{-10mm}
\caption{\emph{Top}: Averaged window for 1 day for the Crab.  \emph{Bottom}: The filtered data with the central portion fit to a polynomial (green), and the outer background portion fit to a spline function (red). }
\end{figure}

\section{Results}
This imaging algorithm has been applied to 17 precession periods of data, \( \sim 2.5 \) years, to generate all-sky images in the 12-25 keV and 100-300 keV energy bands.  The data cover August 12, 2008 to February 12, 2011.  Figure 3 and Figure 4 show the 12-25 keV and 100-300 keV images, respectively.  These are significance maps with green contours at 3.3, 5, 10, and 20\( \sigma \).  The sources in red denote sources previously in the GBM EOT input catalog, while those in blue denote sources added to the input catalog based on the imaging analysis.  These sources were identified through cross-correlating features in the maps with sources in the \emph{Swift}/BAT, \emph{INTEGRAL}/SPI, and \emph{Fermi}/LAT catalogs, which were then analyzed with EOT.  A source is added to the input catalog if the 12-50 keV significance from EOT is at least \( 10 \sigma \).  Table 1 lists the sources found, their flux in the 12-50 keV band, and their statistical significance.  

\begin{table}[t]
\begin{center}
\caption{Sources found with Earth Occultation Imaging}
\begin{tabular}{|l|c|c|}
\hline \textbf{Source Name} & \textbf{Flux (mCrab)} & \textbf{Significance}
\\
\hline 3A 1822-371 & 38.7 & 26.1  \\
\hline 3C 273 & 16.0 & 16.8  \\
\hline LMC X-3 & 16.5 & 14.0  \\
\hline NGC 3227 & 11.1 & 13.4  \\
\hline NGC 5506 & 12.0 & 12.7  \\
\hline IC 4329A & 9.8 & 10.9  \\
\hline IGR J21247+5058 & 13.1 & 10.7  \\
\hline NGC 5252 & 9.8 & 10.7  \\
\hline Coma Cluster & 9.6 & 10.7  \\
\hline
\end{tabular}
\label{l2ea4-t1}
\end{center}
\vspace{-10mm}
\end{table}

\bigskip 


\begin{acknowledgments}
This material is based upon work supported by the Louisiana Optical Network Institute (LONI).  Portions of this research were conducted with high performance computational resources provided by Louisiana State University (http://www.hpc.lsu.edu).  This work is supported by the NASA Fermi Guest Investigator program, NASA/Louisiana Board of Regents Graduate Fellowship Program.  

\end{acknowledgments}

\bigskip 

\newpage

\begin{figure*}[h]
\includegraphics[angle=90,scale=0.5, trim=0mm 40mm 0mm 40mm]{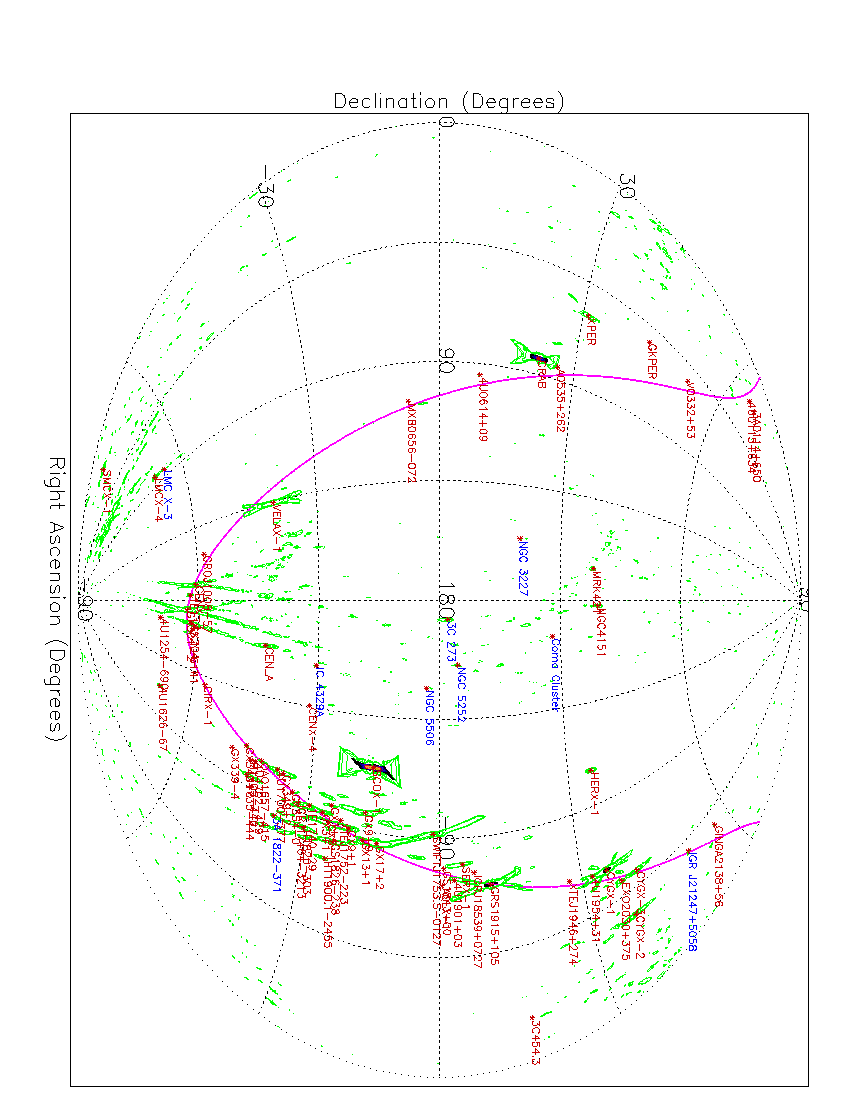}
\caption{All-sky image in the 12-25 keV band for August 12, 2008 to February 12, 2011}
\vspace{-5mm}
\end{figure*}

\begin{figure*}[h]
\includegraphics[angle=90,scale=0.5, trim=0mm 40mm 0mm 40mm]{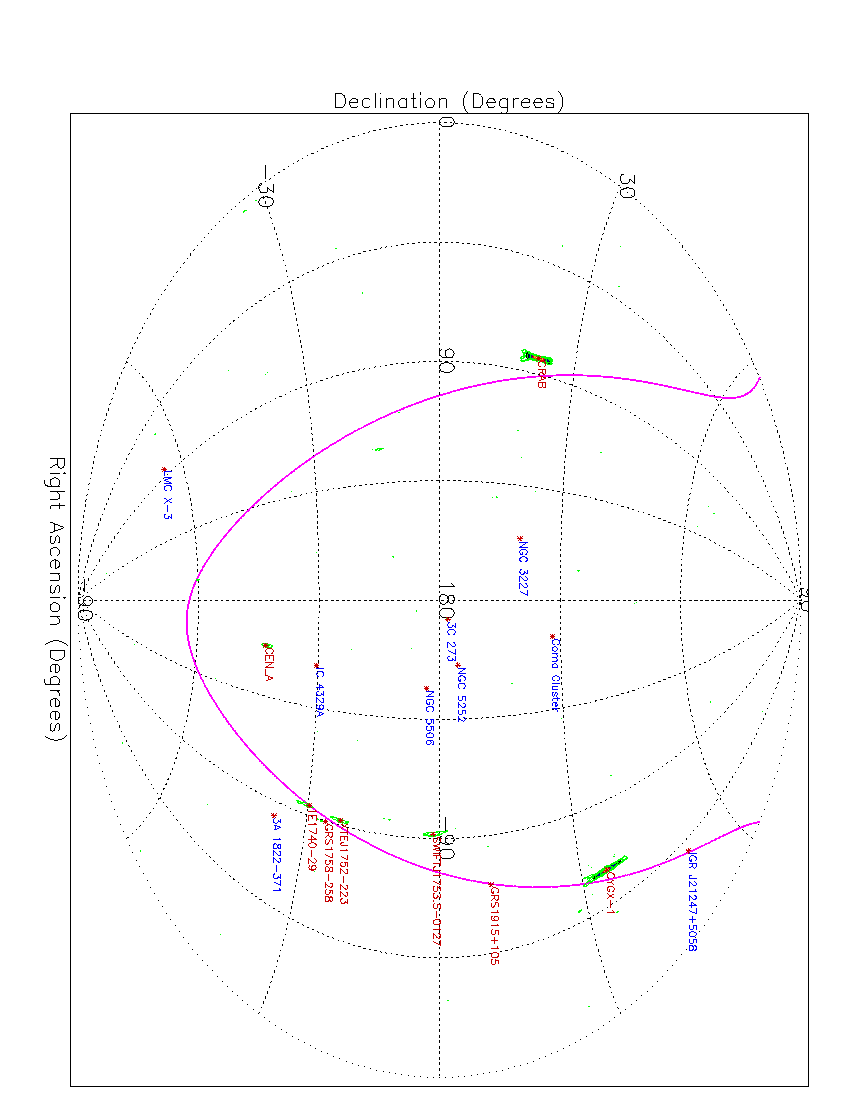}
\caption{All-sky image in the 100-300 keV band for August 12, 2008 to February 12, 2011}
\vspace{-5mm}
\end{figure*}

\end{document}